\documentclass[conf]{new-aiaa}
\usepackage[utf8]{inputenc}

\usepackage{graphicx, bm}
\usepackage{amsmath}
\usepackage{bm}
\usepackage[version=4]{mhchem}
\usepackage{booktabs}
\usepackage{multirow}
\usepackage{siunitx}
\usepackage{longtable,tabularx}
\usepackage{cleveref}
\crefformat{equation}{(#2#1#3)}
\title{Secure Coordination for \\Vertiport Sequencing in Advanced Air Mobility}

\author{Jaehan Im\footnote{Graduate Research Assistant, Department of Aerospace Engineering and Engineering Mechanics, email: jaehan.im@utexas.edu},
Filippos Fotiadis\footnote{Postdoctoral Researcher, Oden Institute for Computational Engineering \& Sciences, email: ffotiadis@utexas.edu},
Ufuk Topcu\footnote{Professor, Oden Institute for Computational Engineering \& Sciences, email: utopcu@utexas.edu}
and
David Fridovich-Keil\footnote{Assistant Professor, Department of Aerospace Engineering and Engineering Mechanics, email: dfk@utexas.edu}}
\affil{The University of Texas at Austin, Austin, TX, 78712}

\begin{document}
\newcommand{\fix}[1]{\textcolor{red}{#1}}

\newcommand{\ones}{\bm 1}
\newcommand{\reals}{{\mbox{\bf R}}}
\newcommand{\integers}{{\mbox{\bf Z}}}
\newcommand{\symm}{{\mbox{\bf S}}}  

\newcommand{\nullspace}{{\mathcal N}}
\newcommand{\range}{{\mathcal R}}
\newcommand{\Rank}{\mathop{\bf Rank}}
\newcommand{\Tr}{\mathop{\bf Tr}}
\newcommand{\diag}{\mathop{\bf diag}}
\newcommand{\card}{\mathop{\bf card}}
\newcommand{\rank}{\mathop{\bf rank}}
\newcommand{\conv}{\mathop{\bf conv}}
\newcommand{\prox}{\bm{prox}}

\newcommand{\Expect}{\mathop{\bf E{}}}
\newcommand{\Prob}{\mathop{\bf Prob}}
\newcommand{\Co}{{\mathop {\bf Co}}} 
\newcommand{\dist}{\mathop{\bf dist{}}}
\newcommand{\argmin}{\mathop{\rm argmin}}
\newcommand{\argmax}{\mathop{\rm argmax}}
\newcommand{\epi}{\mathop{\bf epi}} 
\newcommand{\Vol}{\mathop{\bf vol}}
\newcommand{\dom}{\mathop{\bf dom}} 
\newcommand{\intr}{\mathop{\bf int}}
\newcommand{\sign}{\mathop{\bf sign}}
\newcommand{\norm}[1]{\left\lVert#1\right\rVert}
\newcommand{\mnorm}[1]{{\left\vert\kern-0.25ex\left\vert\kern-0.25ex\left\vert #1 
    \right\vert\kern-0.25ex\right\vert\kern-0.25ex\right\vert}}

\newtheorem{definition}{Definition} 
\newtheorem{theorem}{Theorem}
\newtheorem{lemma}{Lemma}
\newtheorem{corollary}{Corollary}
\newtheorem{remark}{Remark}
\newtheorem{proposition}{Proposition}
\newtheorem{assumption}{Assumption}
\newtheorem{example}{Example}

\newcommand{\cf}{{\it cf.}}
\newcommand{\eg}{{\it e.g.}}
\newcommand{\ie}{{\it i.e.}}
\newcommand{\etc}{{\it etc.}}

\newcommand{\putref}{{\color{red}[r]}}

\newcommand{\ba}[2][]{\todo[color=orange!40,size=\footnotesize,#1]{[BA] #2}}

\newcommand{\bigO}{\mathcal{O}}

\newcommand{\intSet}{\mathbb{Z}}
\newcommand{\realSet}{\mathbb{R}}
\newcommand{\natSet}{\mathbb{N}}
\newcommand{\zeroSet}{\bm{0}}
\newcommand{\state}{\bm{x}}

\newcommand{\param}{\kappa}

\newcommand{\plSet}{\mathbf{N}}
\newcommand{\chSet}{\mathbf{M}}
\newcommand{\opSet}{\mathcal{O}}

\newcommand{\xVec}{\bm{x}}
\newcommand{\coordFactor}{\mathbf{w}}
\maketitle

\begin{abstract}
Advanced air mobility operations will require reliable coordination mechanisms for managing dense traffic near vertiports.
However, sequencing decisions may be vulnerable to falsified self-reported information such as estimated time of arrival.
Self-interested vehicles may misreport their arrival times to obtain favorable landing priority, while malicious actors may spoof information to disrupt sequencing decisions or induce unnecessary congestion.
This paper studies secure coordination for vertiport sequencing under sensing uncertainty.
We consider a coordinator that combines self-reported information with independent surveillance measurements to assign arrival times satisfying separation requirements.
Since surveillance-based estimates are uncertain, falsified reports may remain consistent with the sensing uncertainty region and cannot always be rejected correctly.
We therefore formulate a robust sequencing problem that jointly determines the landing order and arrival times using vehicle-specific trust parameters to balance reliance on reports and uncertain surveillance estimates.
We evaluate the proposed coordination rule under four reporting behaviors: malicious, self-interested, random, and honest.
Monte Carlo experiments show that the proposed rule reduces the mean system-level sequencing cost under malicious reporting by approximately 19--26\% relative to non-robust coordination, while incurring an additional cost of 5--11\% under honest reporting.
These results highlight the asymmetric benefits and costs of robust coordination, even with inaccurate trust parameters.
\end{abstract}

\section{Nomenclature}

{\renewcommand\arraystretch{1.0}
\noindent\begin{longtable*}{@{}l @{\quad=\quad} l@{}}
$N$ & number of vehicles approaching the vertiport \\
$i$ & vehicle index, $i\in\{1,\ldots,N\}$ \\
$\tau_i$ & true estimated time of arrival of vehicle $i$ before sequencing intervention \\
$\hat{\tau}_i$ & reported estimated time of arrival of vehicle $i$ \\
$\tilde{\tau}_i$ & surveillance-based estimate of the arrival time of vehicle $i$ \\
$\delta_i$ & reporting deviation of vehicle $i$ \\
$\xi_i$ & surveillance error for vehicle $i$ \\
$\eta_i$ & possible surveillance error within the uncertainty set \\
$\mathcal{U}_i$ & surveillance uncertainty set for vehicle $i$ \\
$\varepsilon_i$ & surveillance uncertainty bound for vehicle $i$ \\
$\lambda_i$ & coordinator's trust in vehicle $i$'s report \\
$\pi$ & landing order \\
$a_i$ & assigned arrival time of vehicle $i$ after sequencing \\
$d_{\min}$ & minimum required temporal separation between consecutive arrivals \\
$J_i$ & schedule-adjustment cost of vehicle $i$ \\
$J_{\mathrm{sys}}$ & system-level sequencing cost \\
$\mathcal{M}$ & set of vehicles that may misreport their arrival times \\
$S_\lambda$ & sequencing rule parameterized by the trust vector $\lambda$ \\
$r$ & relative surveillance resolution \\
\end{longtable*}}
\setcounter{table}{0}
\section{Introduction}
\vspace{3mm}

\lettrine{A}{dvanced} air mobility (AAM) will require reliable coordination mechanisms for managing dense traffic near vertiports \cite{FAA_UAM_ConOps, federal2023advanced}.
As multiple vehicles approach a shared landing facility, sequencing decisions must be made under limited landing capacity and uncertain arrival-time information \cite{vertiport_capacity}.
Broadcast-based information, such as Remote ID, can support this coordination by providing vehicle identity, position, and other operational data.

Such coordination mechanisms become vulnerable when they rely on self-reported information. 
Self-interested vehicles may strategically manipulate their reported states to obtain more favorable coordination outcomes.
Moreover, because Remote ID is not inherently spoofing-proof, the same coordination infrastructure may also be exposed to adversarial attacks that aim to induce unnecessary congestion or increase collision risk \cite{aiaa_spoofing, Detecting_Misbehave, webb2026remote}. 
While related vulnerabilities are already recognized in conventional air traffic management \cite{comm_to_incid, management_impact, vectoring_to_ineff, ETA_sensitivity_TMA}, we expect them to become more pronounced in AAM operations, where higher traffic density and tighter sequencing margins are expected to increase the operational impact of false information.

This vulnerability of AAM to false reporting is particularly relevant in vertiport sequencing. 
Aircraft sequencing and scheduling have been extensively studied as a means of improving traffic efficiency while satisfying operational and separation constraints \cite{ikli2021aircraft, vertiport_capacity}.
When several vehicles are expected to arrive within a similar time window, even a small change in reported estimated time of arrival can affect the assigned landing order \cite{ETA_sensitivity_TMA, management_impact, vectoring_to_ineff}. 
This order then determines the separation-feasible arrival schedule issued by the coordinator. 
As a result, a vehicle that moves earlier in the sequence may receive a reassigned arrival time closer to its preferred plan, whereas other vehicles may be delayed or required to adjust their approach. 

Coordination problems in air traffic management and AAM, including sequencing and resource allocation, have also been studied from a game-theoretic perspective \cite{shoham2008multiagent}.
In conventional air traffic management, game-theoretic approaches have been employed to address strategic behavior in landing-slot allocation and to design incentive-compatible allocation mechanisms \cite{schummer2017incentives, dixit2023algorithmic}.
Recent studies in air traffic coordination, including AAM, have addressed computational scalability in sequencing problems \cite{rrce, espejo2023heuristic}, robustness to uncertainty in operational information and agents' costs \cite{j_ccce, pang2024chance}, and coordination without requiring agents to disclose their private preferences \cite{taco, schutz2021privacy}.
However, they do not explicitly address the security of coordination against falsified arrival-time information.
While spoofing-aware trajectory planning has been investigated for collision avoidance \cite{webb2026remote}, the impact of false reporting on system-level coordination performance and the associated trade-off between robustness and coordination efficiency remain largely unexplored.

A natural defense against false reporting is to compare self-reported information with independently obtained measurements. In this work, we assume that a coordinator is equipped with an active surveillance system, similar to conventional air traffic management systems \cite{ssr, ssr_error, sspar_error}, that estimates vehicle arrival times independently of self-reported information. However, measurement noise introduces uncertainty into these estimates \cite{ssr_error, sspar_error, van1993phased}, allowing false reports to remain consistent with surveillance measurements. This limited detectability gives rise to a strategic interaction in which vehicles may manipulate their reports to influence sequencing outcomes, while the coordinator must balance its reliance on potentially falsified reports against uncertain surveillance-based estimates.

We investigate secure coordination for AAM vertiport sequencing under sensing uncertainty by considering two sources of intentional false reporting.
The first is \emph{strategic misreporting} by self-interested vehicles, whose objective is to improve their own sequencing outcomes by reducing their schedule-adjustment costs.
The second is \emph{malicious spoofing}, whose objective is not to improve the outcome of a particular vehicle but to degrade the overall sequencing performance.
To address these sources of false reporting, we formulate a robust sequencing problem, solvable through a sequence of convex quadratic programs, that combines self-reported arrival times and surveillance-consistent uncertainty sets.
We evaluate its performance under four reporting behaviors---malicious, self-interested, random, and honest---and investigate how performance depends on the accuracy of the coordinator's trust parameters and surveillance uncertainty.

\section{Vertiport Sequencing Problem} \label{sec:VSP}
\vspace{3mm}

We consider a set of $N$ vehicles approaching a shared vertiport. Each vehicle $i \in \{1,\ldots,N\}$ has a true estimated time of arrival $\tau_i \in \realSet$, which represents the arrival time expected under its current approach plan before sequencing intervention. The coordinator receives a reported arrival time
\begin{equation}
    \hat{\tau}_i = \tau_i + \delta_i,
\end{equation}
where $\delta_i \in \realSet$ denotes the reporting deviation. A truthful report corresponds to $\delta_i = 0$, whereas $\delta_i \neq 0$ represents false reporting. Negative values of $\delta_i$ correspond to reports that claim an earlier arrival time.

The coordinator also receives surveillance-based measurements, which are used to infer an independent estimate of the vehicle's arrival time,
\begin{equation}
    \tilde{\tau}_i = \tau_i + \xi_i,
\end{equation}
where $\xi_i \in \realSet$ denotes the surveillance error.
We associate each vehicle with a surveillance uncertainty set
\begin{equation} \label{eq:uncertaintySet}
    \mathcal{U}_i = [-\varepsilon_i,\varepsilon_i],
\end{equation}
where $\varepsilon_i \in \realSet_{\geq0}$ captures the uncertainty in arrival-time estimation from surveillance data.
We assume that $\xi_i \in \mathcal{U}_i$ with high probability.
Because these measurements are uncertain, false reports may not always be distinguishable from truthful reports.

Given the reported arrival times, the coordinator assigns a landing sequence and constructs a separation-feasible arrival schedule. Let $a_i \in \realSet$ denote the assigned arrival time for vehicle $i$ after sequencing. Each vehicle incurs an adjustment cost $J_i: \realSet \times \realSet \to \realSet$ relative to its true arrival plan,
\begin{equation}
    J_i(a_i,\tau_i) = (a_i-\tau_i)^2.
\end{equation}
This cost captures the operational burden of requiring a vehicle to arrive earlier or later than its original arrival plan.

The system-level sequencing cost $J_{\text{sys}}:\realSet^N \times \realSet^N \to \realSet$ is defined as
\begin{equation} \label{eq:sysCost}
    J_{\mathrm{sys}}(a,\tau)
    = \sum_{i=1}^N J_i(a_i,\tau_i).
\end{equation} 
However, the coordinator does not directly observe $\tau_i$ and instead computes the nominal schedule using the reported arrival time $\hat{\tau}_i$.
Let $\pi$ denote the landing order obtained by sorting the reported arrival times in ascending order, i.e.,
\begin{equation}
    \hat{\tau}_{\pi(1)}
    \leq \cdots \leq
    \hat{\tau}_{\pi(N)}.
\end{equation}
\noindent
The coordinator computes a \textit{nominal} separation-feasible arrival schedule by solving
\begin{equation} \label{eq:NominalCoordProb}
\begin{aligned}
    \min_{a} \quad & \sum_{i=1}^N J_i(a_i,\hat{\tau}_i) \\
    \text{s.t.} \quad
    & a_{\pi(k+1)}-a_{\pi(k)} \geq d_{\min},
    \quad k=1,\ldots,N-1,
\end{aligned}
\end{equation}
where $d_{\min}$ denotes the minimum required temporal separation between consecutive arrivals.
Throughout this work, we assume that all vehicles comply with their assigned arrival times.

\begin{remark}[Schedule compliance]
Removing the assumption of schedule compliance is an important direction for future research. Incentive mechanisms based on monitoring and penalizing deviations from assigned schedules may help induce compliance.
\end{remark}

\section{Secure Coordination via Robust Optimization}
\label{sec:secure_coordination}
\begin{figure}[!htb]
    \centering
    \includegraphics[width=0.9\linewidth]{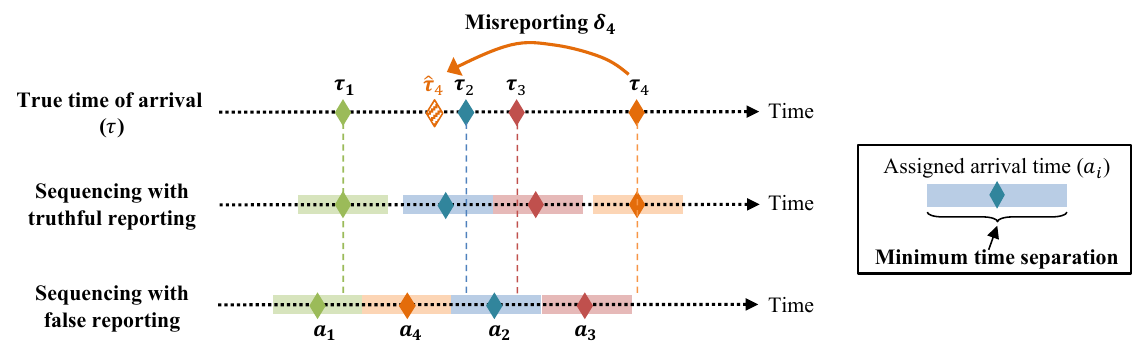}
    \caption{Illustration of the impact of false reporting on vertiport sequencing.
    The top row shows the true arrival times $\tau_i$.
    The second and third rows show the sequencing outcomes under truthful and false reporting, respectively.
    In these two rows, diamonds denote the assigned arrival times $a_i$, and the shaded rectangles represent the corresponding minimum time-separation regions.
    The dashed vertical lines mark the true arrival times to visualize the schedule adjustments.
    By reporting an earlier arrival time, vehicle 4 changes the landing order and shifts the assigned arrival times of all vehicles.
    In this example, false reporting increases the sum of squared deviations from the true arrival times, resulting in a higher system-level sequencing cost as defined in \Cref{eq:sysCost}.}
    \label{fig:spoofing_example}
\end{figure}

\medskip
False reporting or spoofing affects the outcome of \Cref{eq:NominalCoordProb} by altering the reported arrival-time information used to determine the landing order and, consequently, the assigned arrival times.
A false report can therefore alter both an individual vehicle's adjustment cost and the total system-level sequencing cost.
\Cref{fig:spoofing_example} illustrates how malicious reporting can alter the landing order and increase the system-level sequencing cost.

\subsection{Sources of False Information}

We consider two sources of intentional false information in vertiport sequencing.
The first is \emph{self-interested misreporting}.
In this case, a vehicle manipulates its reported estimated time of arrival to improve its own sequencing outcome after observing the coordination rule; for example, it may misreport to obtain an assigned arrival time closer to its original arrival plan, potentially at the expense of other vehicles.

The second source is \emph{malicious spoofing}.
In this case, false information is deliberately introduced with the objective of degrading system-level sequencing performance.
Unlike a self-interested vehicle, a malicious actor is not interested in minimizing the cost of a particular vehicle.
Instead, malicious actors seek reports that produce unfavorable sequencing outcomes, such as increased schedule-adjustment costs or disrupted arrival ordering.

Although these two sources of false information arise from different objectives, the coordinator may not know the underlying objectives of the vehicles.
We therefore seek a robust coordination rule that mitigates the impact of false reporting regardless of the underlying reporting behavior.

We assume that reporting deviations are bounded by the surveillance uncertainty bound, i.e., $|\delta_i|\leq\varepsilon_i$.
Accordingly, the range of feasible reported arrival times is given by
\begin{equation}
    \hat{\tau}_i \in
    [\tau_i-\varepsilon_i,\tau_i+\varepsilon_i],
\end{equation}
where we recall that $\hat{\tau}_i$ denotes the reported arrival time of vehicle $i$.
Since $\varepsilon_i$ is known to the vehicles, larger reporting deviations are more likely to be flagged as inconsistent with surveillance, which motivates the deviation bound.

\subsection{Surveillance-Consistent Uncertainty Sets}
\begin{figure}[b!]
    \centering
    \includegraphics[width=0.65\linewidth]{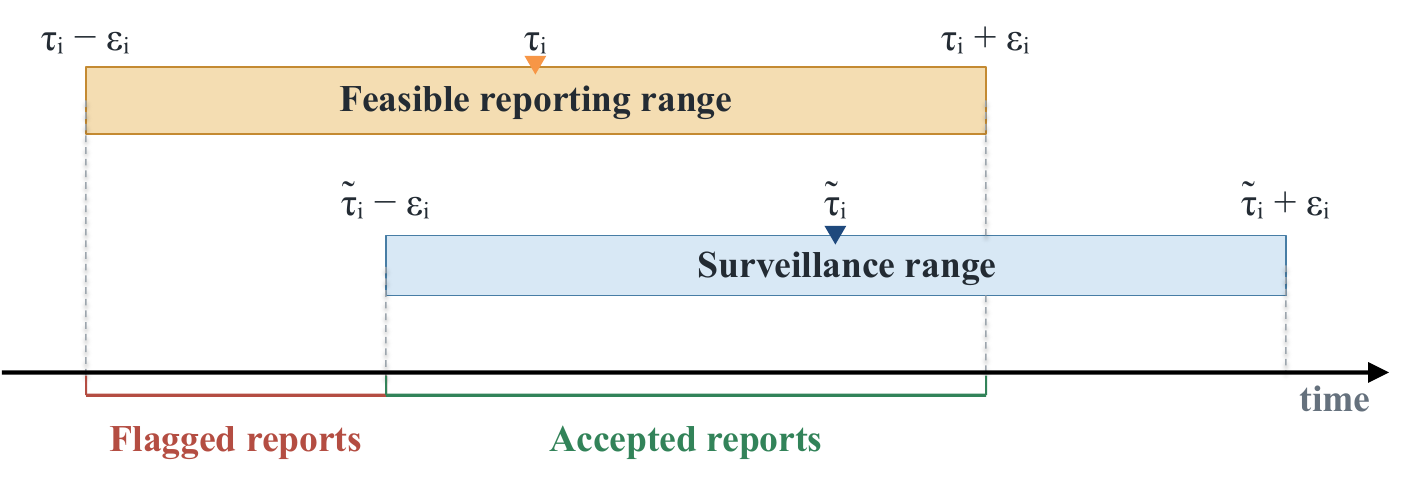}
    \caption{Comparison between feasible reporting range and surveillance range.}
    \label{fig:reportingRange}
\end{figure}

Using the uncertainty-set model introduced in \Cref{sec:VSP}, the coordinator can restrict the range of plausible arrival times for each vehicle.
Specifically, the surveillance-consistent set of possible true arrival times is given by
\begin{equation}
    \tilde{\tau}_i + \mathcal{U}_i
    =
    [\tilde{\tau}_i-\varepsilon_i,
    \tilde{\tau}_i+\varepsilon_i].
\end{equation}
\Cref{fig:reportingRange} illustrates the relationship between the feasible reporting range and the surveillance-consistent uncertainty set.

Reports outside this interval can be flagged as inconsistent with the surveillance-based estimate.
However, reports within this interval cannot be reliably distinguished from truthful reports using the uncertainty bound alone.
Rather than relying solely on the detection of false reports, we explicitly account for the remaining surveillance uncertainty when constructing the arrival schedule through robust optimization.

Even when the reporting deviation is bounded by
$|\delta_i|\leq\varepsilon_i$, the reported arrival time
may fall outside the surveillance-consistent uncertainty set.
In particular, when $\xi_i\in\mathcal{U}_i$,
\begin{equation}
    |\hat{\tau}_i-\tilde{\tau}_i|
    = |\delta_i-\xi_i|
    \leq 2\varepsilon_i.
\end{equation}
Thus, even a bounded reporting deviation may be flagged
due to surveillance error.
In this work, whenever
$|\hat{\tau}_i-\tilde{\tau}_i|>\varepsilon_i$,
the coordinator discards the reported arrival time
and uses the surveillance-based estimate instead.

\begin{remark}[Incentivizing truthful reporting]
Beyond screening, inconsistencies between reported and surveillance-based arrival times may be used to penalize potentially false reports.
This motivates exploring a mechanism design approach to incentivize truthful reporting and enhance coordination security.
\end{remark}

\subsection{Robust Formulation for False Reporting}

To account for different levels of confidence in the reported information, we associate each vehicle with a trust parameter $\lambda_i\in[0,1]$.
A larger $\lambda_i$ places greater emphasis on the reported arrival time, whereas a smaller $\lambda_i$ places greater emphasis on the surveillance-based estimate and its associated uncertainty.
For each vehicle whose report is flagged as inconsistent with surveillance, we set $\lambda_i=0$ and replace $\hat{\tau}_i$ with $\tilde{\tau}_i$.

The coordinator jointly determines the landing order and assigned arrival times by solving the following robust sequencing problem:
\begin{equation}
\label{eq:RobustCoordProb}
\begin{aligned}
    \min_{a}\;
    \max_{\substack{\eta_i\in\mathcal{U}_i,\\ i=1,\ldots,N}}
    \quad &
    \sum_{i=1}^{N}
    \left[
        (1-\lambda_i)
        J_i(a_i,\tilde{\tau}_i-\eta_i)
        +
        \lambda_i
        J_i(a_i,\hat{\tau}_i)
    \right]
    \\
    \text{s.t.}\quad
    &
    |a_i-a_j|\geq d_{\min},
    \quad \forall i\neq j.
\end{aligned}
\end{equation}
Here, $\eta_i$ represents a possible realization of the surveillance error within $\mathcal{U}_i$.
The uncertainty sets and trust parameters are given to the coordinator, and the landing order is determined by the assigned arrival-time vector $a$.

Since the objective is separable with respect to $\eta_i$, the inner maximization admits a closed-form solution,
\begin{equation}
    \max_{\eta_i\in\mathcal{U}_i}
    J_i(a_i,\tilde{\tau}_i-\eta_i)
    =
    \left(|a_i-\tilde{\tau}_i|+\varepsilon_i\right)^2.
\end{equation}
For any fixed landing order $\pi$, \Cref{eq:RobustCoordProb} reduces to the following convex quadratic program:
\begin{equation}
\label{eq:RobustCoordQP}
\begin{aligned}
    \min_{a,t}\quad &
    \sum_{i=1}^{N}
    \left[
        (1-\lambda_i)(t_i+\varepsilon_i)^2
        +
        \lambda_i(a_i-\hat{\tau}_i)^2
    \right]
    \\
    \text{s.t.}\quad
    &
    t_i\geq a_i-\tilde{\tau}_i,
    \quad i=1,\ldots,N,
    \\
    &
    t_i\geq \tilde{\tau}_i-a_i,
    \quad i=1,\ldots,N,
    \\
    &
    a_{\pi(k+1)}-a_{\pi(k)}
    \geq d_{\min},
    \quad k=1,\ldots,N-1,
\end{aligned}
\end{equation}
where $t$ is a vector of auxiliary variables introduced to express the absolute values in the objective function \cite{boyd}.
A globally optimal schedule can therefore be obtained by selecting the solution with the lowest objective value across all possible landing orders.

When $\lambda_i=1$ for all vehicles and report screening is omitted, the robust sequencing problem reduces to the nominal coordination problem in \Cref{eq:NominalCoordProb}.
Conversely, when $\lambda_i=0$, the coordinator disregards the report from that vehicle and determines the assigned arrival time based on the worst-case adjustment cost over its surveillance uncertainty set.

\section{Numerical Experiments}
\label{sec:experiments}

We conduct Monte Carlo experiments to evaluate the performance of the proposed robust coordination rule under different reporting behaviors, levels of report trust, and surveillance uncertainty.

\subsection{Experimental Setup}

We consider five vehicles ($N=5$) approaching a single vertiport.
The true arrival times are independently sampled from $\tau_i\sim\mathrm{U}(0,10)$, and the minimum required temporal separation is set to $d_{\min}=2$.
We assume a homogeneous surveillance system (\ie, $\varepsilon_i=\varepsilon$ for all $i$) and sample $\varepsilon\sim\mathrm{U}(0.25,1.25)$ for each scenario.
The surveillance errors are independently sampled from
\begin{equation}
    \xi_i\sim\mathcal{N}
    \left(0,\left(\frac{\varepsilon}{2.576}\right)^2\right),
\end{equation}
such that $\mathrm{P}(\xi_i\in\mathcal{U}_i)\approx 0.99$ for each vehicle, where we recall that $\mathcal{U}_i=[-\varepsilon_i,\varepsilon_i]$ denotes the surveillance uncertainty set of vehicle $i$.

We let $\mathcal{M}=\{2,4\}$ denote the set of vehicles that may misreport their arrival times. All other vehicles report truthfully.
Thus, we impose $|\delta_i|\leq\varepsilon$ for all $i\in\mathcal{M}$ and $\delta_i=0$ otherwise.
We generate 1,000 independent scenarios and evaluate all reporting behaviors and coordinator configurations on the same set of scenarios.

\subsection{Reporting Behaviors}

We consider four reporting behaviors for the vehicles in $\mathcal{M}$.
Let $S_{\lambda}:\mathbb{R}^{N}\times\mathbb{R}^{N}\to\mathbb{R}^{N}$ denote a sequencing rule with report screening that maps the vectors of reported and surveillance-based arrival times, $\hat{\tau}$ and $\tilde{\tau}$, to the assigned arrival-time vector $a$.
We denote its $i$th component by $S_{\lambda,i}$.

\medskip
\begin{enumerate}
    \item \textbf{Malicious}: Vehicles in $\mathcal{M}$ jointly select their reporting deviations to maximize the system-level sequencing cost, while all other vehicles report truthfully ($\delta_i=0$ for all $i\notin\mathcal{M}$):
    \begin{equation}
        \delta_{\mathcal{M}}^{\mathrm{mal}}
        \in
        \arg\max_{\substack{|\delta_i|\leq\varepsilon\\
        i\in\mathcal{M}}}
        J_{\mathrm{sys}}\left(
            S_{\lambda}(\tau+\delta,\tilde{\tau}),\tau
        \right).
    \end{equation}
    \item \textbf{Self-interested}: Each vehicle $i\in\mathcal{M}$ independently selects its reporting deviation to minimize its own sequencing cost, assuming that all other vehicles report truthfully:
    \begin{equation}
        \delta_i^{\mathrm{self}}
        \in
        \arg\min_{|\delta_i|\leq\varepsilon}
        J_i\left(
            S_{\lambda,i}(\tau+\delta_i e_i,\tilde{\tau}),\tau_i
        \right),
    \end{equation}
    where $e_i$ denotes the $i$-th standard basis vector.
    This represents myopic decision-making, with all vehicles submitting their reports simultaneously.
    
    \item \textbf{Random}:
    Each vehicle $i\in\mathcal{M}$ independently samples its reporting deviation from $\delta_i\sim\mathrm{U}(-\varepsilon,\varepsilon)$.

    \item \textbf{Honest}: 
    All vehicles report truthfully (\ie, $\delta_i=0$ for all $i$).
\end{enumerate}
\medskip
For malicious and self-interested reporting, we assume that the vehicles employ some decision-making mechanism to determine their reporting deviations, without specifying how they obtain or estimate the information required for such decisions. 
To evaluate the robustness of the proposed coordination rule against well-informed vehicles, we assume that these vehicles have exact knowledge of $\tau$, $\tilde{\tau}$, $\varepsilon$, $\lambda$, and the coordination rule, including the report screening mechanism.

\subsection{Coordinator Configurations}

We consider four coordinator configurations with different trust parameters, as summarized in \Cref{tab:coordinators}.

\begin{table}[htbp]
    \centering
    \caption{Trust parameters of the coordinator configurations.}
    \label{tab:coordinators}
    \begin{tabular}{lcc}
        \toprule
        Coordinator & $\lambda$ & $\|\Delta\lambda\|_2$ \\
        \midrule
        \texttt{Oracle}
        & $[1,0,1,0,1]$ & 0.00 \\
        \texttt{Imprecise-0.3}
        & $[0.7,0.3,0.7,0.3,0.7]$ & 0.67 \\
        \texttt{Imprecise-0.6}
        & $[0.4,0.6,0.4,0.6,0.4]$ & 1.34 \\
        \texttt{Non-robust}
        & $[1,1,1,1,1]$ & 1.41 \\
        \bottomrule
    \end{tabular}
\end{table}

The \texttt{Oracle} coordinator has perfect knowledge of $\mathcal{M}$.
The \texttt{Imprecise-0.3} and \texttt{Imprecise-0.6} coordinators represent different levels of inaccuracy in this knowledge.
We quantify the discrepancy in the trust parameters as
\begin{equation}
    \|\Delta\lambda\|_2
    =
    \|\lambda-\lambda^{\mathrm{\texttt{Oracle}}}\|_2.
\end{equation}
The \texttt{Non-robust} coordinator serves as the baseline and solves the nominal coordination problem in \Cref{eq:NominalCoordProb}.
All coordinator configurations except the \texttt{Non-robust} coordinator perform report screening before solving the sequencing problem.
The suffixes 0.3 and 0.6 indicate the magnitude of the deviation of each trust parameter from its \texttt{Oracle} counterpart.

\subsection{Performance Evaluation}

We evaluate the performance of each coordinator using the system-level sequencing cost $J_{\mathrm{sys}}(a,\tau)$ computed with respect to the true arrival times.
We consider the following three performance metrics.

\begin{enumerate}

    \item \textbf{Paired difference in system-level costs}: To compare the performance of different coordinators while accounting for scenario-specific variability, we define
    \begin{equation}
        \Delta J_{\mathrm{sys}}
        = J_{\mathrm{sys}}(a,\tau)
        - J_{\mathrm{sys}}(a^{\mathrm{NR}},\tau),
    \end{equation}
    where $a^{\mathrm{NR}}$ denotes the schedule obtained by the \texttt{Non-robust} coordinator under the same scenario and reporting behavior.
    A negative difference indicates an improvement over non-robust coordination.

    \item \textbf{Conditional value-at-risk (CVaR):}
    In addition to the mean cost, we use the conditional value-at-risk at the 95\% confidence level to evaluate performance under adverse scenarios.
    The empirical $\mathrm{CVaR}_{0.95}$ is computed as the average cost over the 5\% most costly scenarios.

    \item \textbf{Gini index:}
    To evaluate the fairness of coordination outcomes, we use the Gini index as in \cite{rrce},
    \begin{equation}
        G(a,\tau)
        =
        \frac{1}{2N J_{\mathrm{sys}}(a,\tau)}
        \sum_{i=1}^{N}\sum_{j=1}^{N}
        \left|J_i(a_i,\tau_i)-J_j(a_j,\tau_j)\right|.
        \label{eq:gini_index}
    \end{equation}
    A Gini index of zero indicates an equal allocation of schedule-adjustment costs across vehicles, whereas larger values indicate greater inequality.
    We set \(G(a,\tau)=0\) when \(J_{\mathrm{sys}}(a,\tau)=0\).

\end{enumerate}

\subsection{Results}

\subsubsection{Coordination efficiency}

\begin{figure}[b!]
    \centering
    \includegraphics[width=0.7\linewidth]{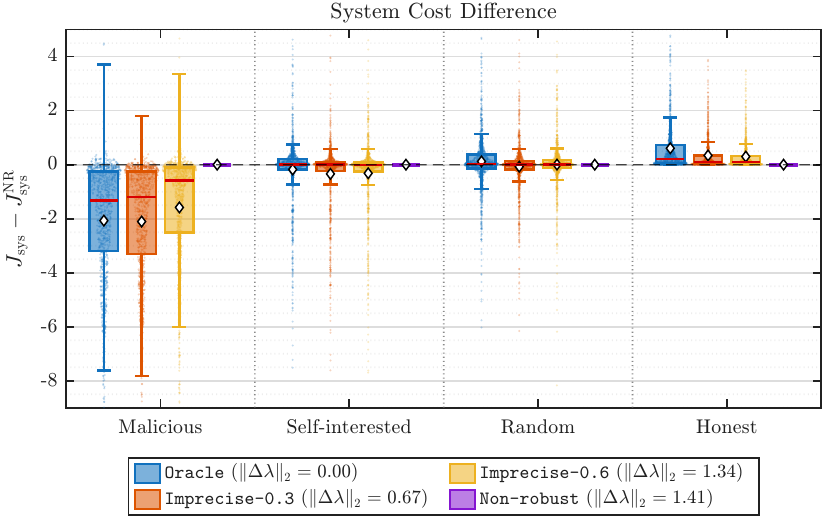}
    \caption{Paired differences in system-level costs relative to the \texttt{Non-robust} coordinator under different reporting behaviors and coordinator configurations. The boxes represent the interquartile ranges, the red lines indicate the medians, and the diamonds ($\diamond$) indicate the mean differences. Each dot represents an individual scenario.}
    \label{fig:normalizedCost}
\end{figure}

\Cref{fig:normalizedCost} presents the differences in system-level sequencing costs relative to the \texttt{Non-robust} coordinator under the four reporting behaviors.
Under malicious reporting, the three robust coordinators achieve approximately 19--26\% lower mean system-level costs than the \texttt{Non-robust} coordinator.
Substantial cost reductions are maintained even with inaccurate trust parameters.

Under self-interested reporting, the robust coordinators reduce the mean system-level cost by approximately 3--5\% relative to the \texttt{Non-robust} coordinator.
Under random reporting, the mean system-level costs remain within approximately 2\% of the non-robust baseline.
Under honest reporting, however, robust coordination increases the mean system-level cost by approximately 11\% for the \texttt{Oracle} coordinator and 5--6\% for the two imprecise coordinators.

\subsubsection{Benefit and cost of robust coordination}

\begin{table}[t!]
    \centering
    \caption{Mean and conditional value-at-risk of system-level sequencing costs under malicious and honest reporting. Values in parentheses indicate the cost difference relative to the \texttt{Non-robust} coordinator.}
    \label{tab:tailRisk}
    \begin{tabular}{lcccc}
        \toprule
        & \multicolumn{2}{c}{Mean cost}
        & \multicolumn{2}{c}{$\mathrm{CVaR}_{0.95}$} \\
        \cmidrule(lr){2-3}
        \cmidrule(lr){4-5}
        Coordinator
        & Malicious & Honest
        & Malicious & Honest \\
        \midrule
        \texttt{Oracle}
        & 6.11 ($-2.07$) & 6.11 ($+0.61$)
        & 20.67 ($-4.94$) & 20.67 ($+1.66$) \\
        \texttt{Imprecise-0.3}
        & 6.08 ($-2.11$) & 5.85 ($+0.34$)
        & 20.48 ($-5.14$) & 19.85 ($+0.83$) \\
        \texttt{Imprecise-0.6}
        & 6.60 ($-1.58$) & 5.80 ($+0.30$)
        & 21.82 ($-3.80$) & 19.43 ($+0.42$) \\
        \texttt{Non-robust}
        & 8.18 & 5.50
        & 25.62 & 19.02 \\
        \bottomrule
    \end{tabular}
\end{table}

\Cref{tab:tailRisk} summarizes the mean and conditional value-at-risk of the system-level sequencing costs under malicious and honest reporting.
The asymmetric benefits and costs of robust coordination are observed in both the mean cost and the upper tail of the cost distribution.

Over the 5\% most costly scenarios ($\mathrm{CVaR}_{0.95}$), robust coordination reduces the system-level cost under malicious reporting by approximately 15--20\%, while increasing it under honest reporting by 2--9\%.
In absolute terms, the benefit is approximately 3--9 times greater than the cost, compared with a factor of 3--6 for the mean cost.

These results suggest that the benefits of robust coordination outweigh its additional costs not only on average but also when considering high-cost scenarios.

\begin{figure}[b!]
    \centering
    \includegraphics[width=0.7\linewidth]{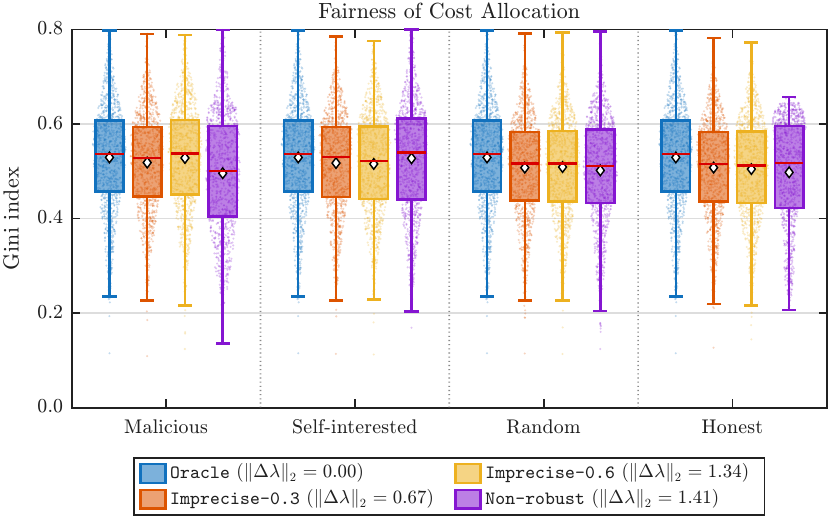}
    \caption{Gini indices of schedule-adjustment costs across vehicles under different reporting behaviors and coordinator configurations.
    The boxes represent the interquartile ranges, the red lines indicate the medians, and the diamonds indicate the mean values.
    Each dot represents an individual scenario.}
    \label{fig:gini}
\end{figure}

\subsubsection{Fairness of robust coordination}

\Cref{fig:gini} presents the distributions of Gini indices under different reporting behaviors and coordinator configurations.
Across all reporting behaviors, the distributions are largely comparable among the four coordinator configurations, with mean Gini indices ranging approximately from 0.50 to 0.53.
This suggests that the benefits of robust coordination can be achieved with limited impact on fairness.

Under malicious reporting, the robust coordinators exhibit slightly higher Gini indices on average than the \texttt{Non-robust} coordinator.
Under the other reporting behaviors, no consistent change in fairness is observed.

\subsubsection{Illustrative Scenario}
\begin{figure}[t!]
    \centering
    \includegraphics[width=0.8\linewidth]{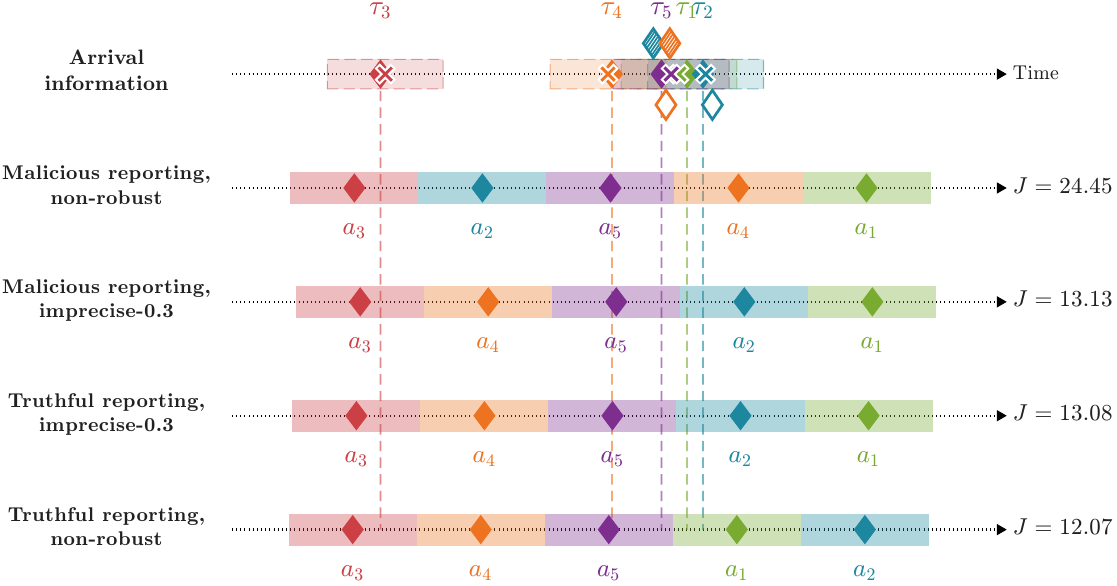}
    \caption{Illustrative scenario constructed using numerical results from a selected Monte Carlo trial. The top row shows the true arrival times $\tau_i$, the surveillance-based estimate $\tilde{\tau}_i$ ($\times$ marker) and the corresponding screening interval $\tilde{\tau}_i \pm \varepsilon$ (dashed shaded rectangle). The hatched diamonds denote the malicious reports chosen against the \texttt{Non-robust} coordinator, and the hollow diamonds denote the malicious reports chosen against the robust (\texttt{Imprecise-0.3}) coordinator. Each subsequent row shows the assigned arrival times $a_i$ under a different coordinator and reporting behavior. In these rows, diamonds denote the assigned arrival times, and the shaded rectangles of width $d_{\min}$ represent the minimum required temporal separation.}
    \label{fig:qualitativeCase}
\end{figure}

\Cref{fig:qualitativeCase} illustrates a representative Monte Carlo scenario in which malicious reporting affects the \texttt{Non-robust} coordinator but has little effect on the \texttt{Imprecise-0.3} coordinator. 
Under the \texttt{Non-robust} coordinator, vehicle 2 reports an earlier arrival time and vehicle 4 reports a later arrival time, which changes the landing order and increases the system-level sequencing cost from 12.07 to 24.45. 
In contrast, under the robust (\texttt{Imprecise-0.3}) coordinator, the malicious intent causes a cost increase from 13.08 to 13.13, and does not alter the landing order. This example shows that robust coordination can substantially mitigate the impact of malicious reporting.

\subsubsection{The role of surveillance resolution $\varepsilon$}

\begin{figure}[t!]
    \centering
    \includegraphics[width=0.75\linewidth]{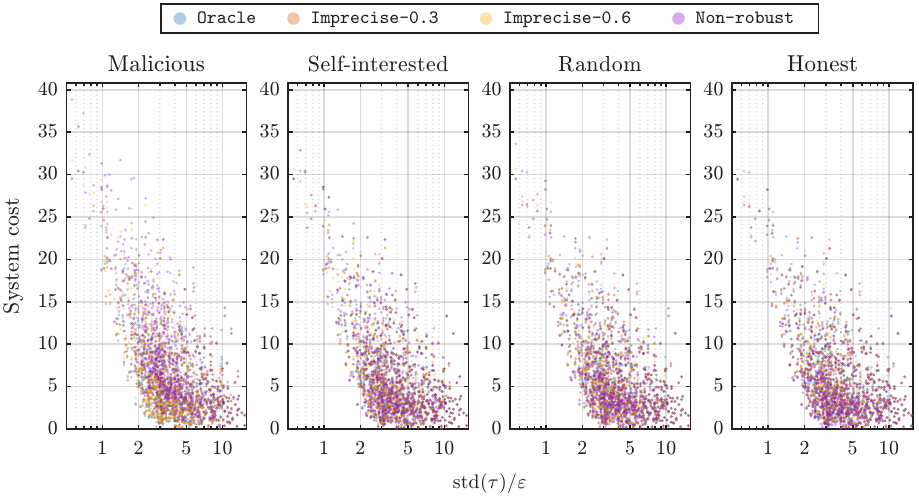}
    \caption{System-level sequencing costs as a function of relative surveillance resolution, $\operatorname{std}(\tau)/\varepsilon$, under different reporting behaviors.
    Each point represents an individual scenario, and the horizontal axis is logarithmically scaled.}
    \label{fig:resolution}
\end{figure}

To further investigate how coordination performance depends on surveillance accuracy relative to the temporal distribution of arriving vehicles, we define a dimensionless metric, named the \emph{relative surveillance resolution} $r$, as
\begin{equation}
    r=\frac{\operatorname{std}(\tau)}{\varepsilon},
\end{equation}
where $\operatorname{std}(\tau)$ denotes the sample standard deviation of the true arrival times.
A larger $r$ indicates finer surveillance resolution relative to the temporal dispersion of vehicles.
We examine how the system-level cost varies with $r$.

\Cref{fig:resolution} presents the system-level sequencing costs as a function of the relative surveillance resolution.
Across all reporting behaviors and coordinator configurations, the system-level costs generally decrease as the relative surveillance resolution increases.

This trend suggests that finer relative surveillance resolution is associated with greater coordination efficiency.
Interestingly, no data points are observed in the region where both the relative surveillance resolution and the system-level cost are low.
The lower envelope of the cost distributions suggests the existence of a lower bound on the achievable system-level cost as a function of the relative surveillance resolution.

\section{Discussion}
\label{sec:discussion}

\subsection{Benefit and cost of robust coordination}

The results in \Cref{fig:normalizedCost} demonstrate a trade-off between mitigating the impact of false reporting and maintaining coordination efficiency under truthful reporting.
As formulated in \Cref{eq:RobustCoordProb}, the proposed coordination rule reduces its reliance on potentially false reports by incorporating the worst-case sequencing cost over the surveillance uncertainty sets.
This provides protection against malicious reporting, but may introduce unnecessary conservatism when the reported information is accurate.
Under random reporting, these two effects appear to offset each other, resulting in little change in system-level cost.

This tradeoff raises the question of when robust coordination is justified.
To illustrate the implications of this asymmetry, suppose that malicious and honest reporting occur with probabilities $p$ and $1-p$, respectively.
Let $\Delta J_{\mathrm{mal}}$ denote the mean cost reduction under malicious reporting and $\Delta J_{\mathrm{hon}}$ denote the mean cost increase under honest reporting, relative to the \texttt{Non-robust} coordinator.
The expected cost reduction achieved by robust coordination is then
\begin{equation}
    p\Delta J_{\mathrm{mal}}-(1-p)\Delta J_{\mathrm{hon}}.
\end{equation}
Thus, robust coordination yields a lower expected system-level cost than non-robust coordination when
\begin{equation}
    p\Delta J_{\mathrm{mal}} >
    (1-p)\Delta J_{\mathrm{hon}}.
\end{equation}
Based on the observed mean costs in \Cref{tab:tailRisk}, the thresholds for $p$ are approximately 23\%, 14\%, and 16\% for the \texttt{Oracle}, \texttt{Imprecise-0.3}, and \texttt{Imprecise-0.6} coordinators, respectively.
This result illustrates how the relatively small cost of robust coordination under honest reporting can be justified by its benefits under malicious reporting.

\subsection{Sensitivity to trust parameters $\lambda$}

The results in \Cref{fig:normalizedCost} show that the performance of robust coordination under malicious reporting depends on the accuracy of the trust parameters.
While \texttt{Imprecise-0.3} achieves performance comparable to that of the \texttt{Oracle} coordinator, \texttt{Imprecise-0.6} achieves a smaller but still substantial reduction in the mean system-level cost.

This observation suggests that accurate identification of potentially false-reporting vehicles may not be necessary to obtain the benefits of robust coordination.
Instead, incorporating independent surveillance information, even with imprecise trust parameters, can mitigate the impact of malicious reporting.

Furthermore, the \texttt{Oracle} coordinator does not necessarily outperform the imprecise coordinators.
Under honest reporting, assigning zero trust to the vehicles in $\mathcal{M}$ results in greater reliance on uncertain surveillance information, leading to higher sequencing costs.
Thus, perfect knowledge of which vehicles may misreport does not necessarily translate into better sequencing performance across all reporting behaviors.

\subsection{Effect of surveillance resolution}

The results in \Cref{fig:resolution} highlight the importance of considering surveillance uncertainty relative to the temporal distribution of arriving vehicles.

When arrival times are concentrated, the minimum separation constraints in \Cref{eq:RobustCoordProb} may require substantial adjustments to the original arrival plans, regardless of the reporting behavior.
When surveillance uncertainty is large, the admissible reporting deviations are also large, increasing the potential impact of false reporting.
Both factors contribute to the higher costs observed at low relative surveillance resolutions.

Robust coordination is therefore particularly beneficial when false reporting can substantially affect sequencing decisions while sufficient flexibility remains in the arrival schedule.
These observations suggest that surveillance resolution and robust coordination play complementary roles in mitigating the impact of false reporting.

\subsection{Robustness and fairness}

The results in \Cref{fig:gini} suggest that robust coordination does not substantially compromise fairness.
Even under malicious reporting, the mean Gini index increases by approximately 4.5--6.7\% relative to the \texttt{Non-robust} coordinator, despite a 19--26\% reduction in the mean system-level cost.
These results suggest that robustness can be improved with a limited impact on fairness.

\section{Conclusion}
\medskip
We investigated secure coordination for vertiport sequencing under false reporting and surveillance uncertainty.
We formulated a robust sequencing problem that jointly determines the landing order and assigned arrival times by combining self-reported arrival times and independent surveillance-based estimates.
For any fixed landing order, it admits a convex quadratic programming reformulation.
The proposed coordination rule accounts for surveillance uncertainty without requiring knowledge of the underlying reporting behavior.

Through Monte Carlo experiments, we demonstrated that robust coordination reduces the mean system-level sequencing cost under malicious reporting by approximately 19--26\%, while incurring an additional cost of 5--11\% under honest reporting.
The benefits of robust coordination were maintained even when the coordinator had inaccurate knowledge of which vehicles may misreport.
Furthermore, the results highlight the importance of surveillance resolution relative to the temporal distribution of arriving vehicles.
These findings suggest that robust coordination can mitigate the impact of false reporting at a relatively small cost to nominal coordination efficiency.

Several directions remain for future work.
These include investigating strategic reporting behaviors under different information structures and designing coordination mechanisms that incentivize trustworthy behavior, including truthful reporting and compliance with assigned arrival schedules.

\section{Acknowledgements}
\medskip
This work was supported by the National Science Foundation CAREER award under grant 2336840, by the National Aeronautics and Space Administration ULI Award under grants 80NSSC21M0071 and 80NSSC24M0070, and by the Office of Naval Research under grant N00014-22-1-2703.
The authors used a large language model to assist with editing and checking mathematical derivations. They confirm that all statements accurately reflect their intended meaning.

\bibliography{sample}

\end{document}